\documentclass[12pt]{article}
\usepackage{graphicx}
\usepackage{siunitx}

\newcommand\pubnumber{NuPhys2017-VanDePontseele}
\newcommand\pubdate{\today}

\def\Title#1{\begin{center} {\Large #1 } \end{center}}
\def\Author#1{\begin{center}{ \sc #1} \end{center}}
\def\Address#1{\begin{center}{ \it #1} \end{center}}

\newcommand\pubblock{\rightline{\begin{tabular}{l} \pubnumber\\
         \pubdate  \end{tabular}}}
\newenvironment{Abstract}{\begin{quotation}  }{\end{quotation}}
\newenvironment{Presented}{\begin{quotation} \begin{center} 
             PRESENTED AT\end{center}\bigskip 
      \begin{center}\begin{large}}{\end{large}\end{center} \end{quotation}}

\begin{document}

\begin{titlepage}
\pubblock

\vfill
\Title{Flash-matching for the $\nu_e$ selection in MicroBooNE}
\vfill
\Author{Wouter Van De Pontseele \\ representing the MicroBooNE collaboration}
\Address{Department of Physics \\ University of Oxford, Oxford, UK \\ Harvard University, Cambridge, MA, USA}
\vfill
\begin{Abstract}
A description of the combination of time projection chamber read-out with optical information in the MicroBooNE detector to reduce cosmogenic backgrounds and select electron neutrino induced interactions.
\end{Abstract}
\vfill
\begin{Presented}
NuPhys2017, Prospects in Neutrino Physics\\
Barbican Centre, London, UK, December 20-22, 2017
\end{Presented}
\vfill
\end{titlepage}
\def\thefootnote{\fnsymbol{footnote}}
\setcounter{footnote}{0}

\section{Introduction}

MircoBooNE is the first liquid argon time projection chamber in an extensive short-baseline neutrino oscillation programme using the booster neutrino beam at Fermilab~\cite{detector}.

The main active element of MicroBooNE is a time projection chamber filled with liquid argon at \SI{87}{\kelvin}. Its dimensions are \SI{10}{\m} in the beam direction, a height of \SI{2.3}{m}, and \SI{2.5}{\m} width. A charged particle that traverses the active volume will interact with the argon in two ways; it will create prompt scintillation light in the ultraviolet region and it will leave a trail of ionization electrons.

Between the two sides of the detector, a high voltage of \SI{70}{\kilo \V} gives rise to a uniform electric field with a magnitude of \SI{273}{\V / \cm}.  The electrons created along the charged particle trajectory will drift towards the three anode wire planes. Behind the wire planes, 32 PMTs are positioned to detect the scintillation light.
This proceeding describes the first efforts towards using the optical system to select electron neutrino's.

\section{Event reconstruction}\label{sec:optical_pre_cuts}

Since MicroBooNE is a surface detector it is subject to an estimated rate of cosmic muons of \SI{5.5}{\Hz}. These cosmic muons are the main source of background to neutrino events. With this in mind, the Panndora reconstruction framework was developed~\cite{pandora}. 

Pandora facilitates the implementation of pattern-recognition algorithms while promoting a multi-algorithm approach in which individual algorithms each address a specific task in a particular topology. 
The waveforms on the wires during the \SI{4.8}{\ms} readout window are taken as input of the reconstruction. 
First, a Gaussian distribution is fitted to each peak in the waveforms, called a \textit{2D hit}.

\begin{figure}[htb]
\centering 
\includegraphics[width= 0.5\textwidth]{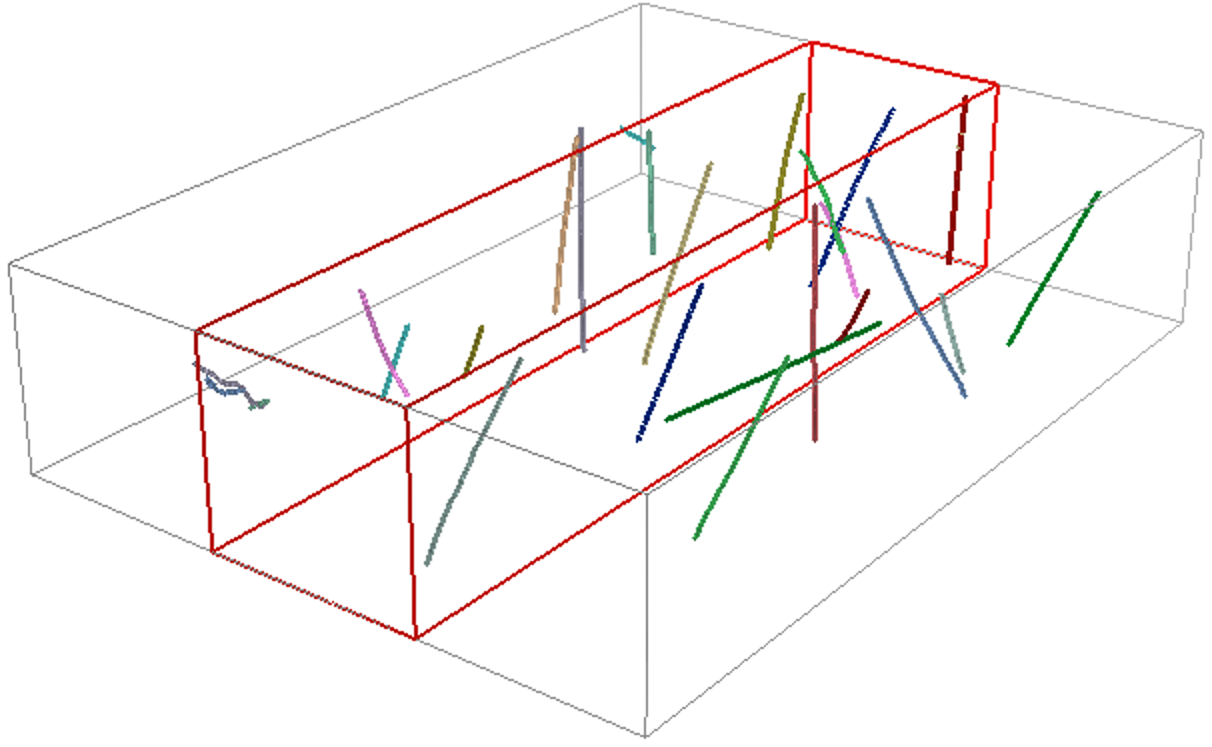} \hfill
\includegraphics[width= 0.45\textwidth]{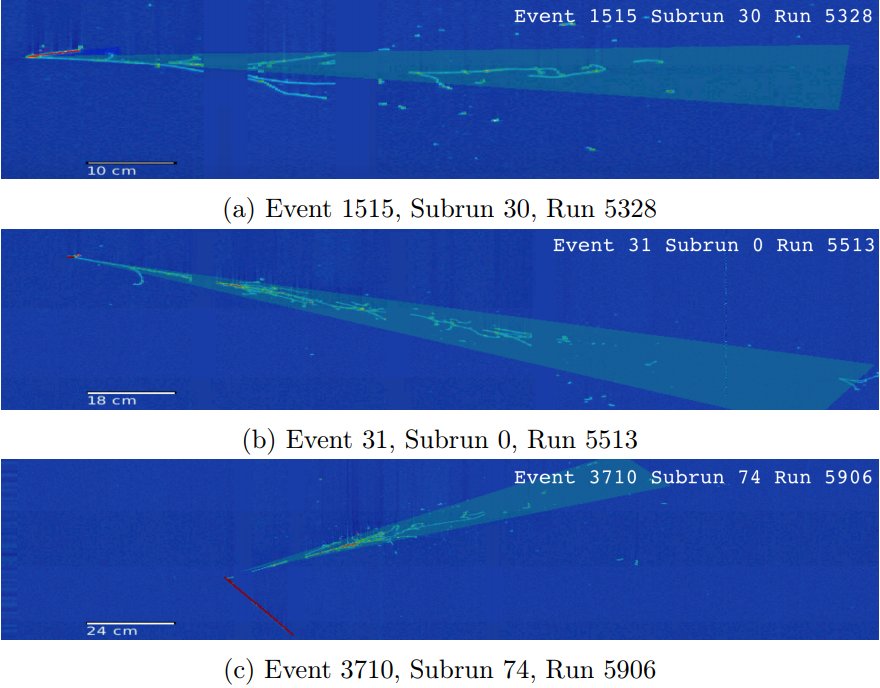} 

\caption{Left: Cosmic reconstruction in the MicrobooNE detector during a readout window of \SI{4.8}{\ms}. Right: The Pandora reconstruction framework groups the remaining objects after cosmic removal together in neutrino candidates and classifies them as tracks and showers. Three selected data events are shown.} 

\label{fig:reco}
\end{figure}

Two subsequent reconstruction paths have been created for use in the MicroBooNE analysis:
\begin{itemize}
\item \textit{PandoraCosmic} is a strongly track oriented selection and aims at unambiguously tagging cosmic-ray muons. Afterwards a cosmic-removed hit collection is created. See the left panel in Figure~\ref{fig:reco}.
\item \textit{PandoraNu} identifies a neutrino interaction vertex and uses it to aid the reconstruction
of all particles, tracks and showers, emerging from the vertex position. A parent neutrino particle is made and the reconstructed particles are added as daughters (Right panel in Figure~\ref{fig:reco}).
\end{itemize}
After this stage, all bits of activity inside the TPC after the cosmic rejection are grouped into Pandora neutrino candidates. Each event contains a number of Pandora neutrino candidates, consisting each of tracks and/or showers.

\begin{figure}[htb]
\centering 
\includegraphics[width= 1.0\textwidth]{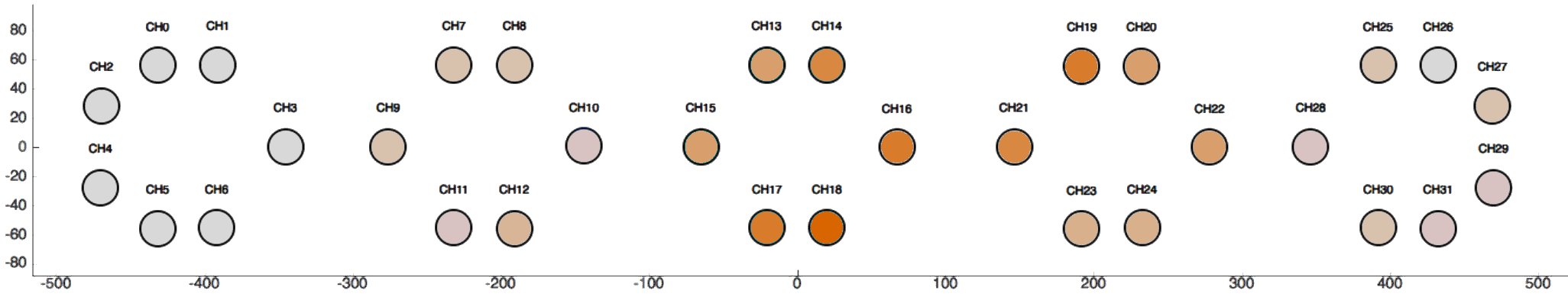}
\caption{An example of an optical reconstructed flash object as seen by the PMT system. Darker orange corresponds to higher photo-electron count.} 
\label{fig:flash}
\end{figure}

Optical hits of the 32 PMTs with the same arrival time withing the event are clustered into flashes. These reconstructed flash objects represent interactions in the detector. It is now required that at least one of these flashes has a timestamp within the window corresponding to the \SI{1.6}{\us} duration of a BNB spill. The flash is required to consist of at least 50 photo-electrons (PE).

\section{Optical Flash Matching}

At this point it is guaranteed that the event has a properly reconstructed flash. A flash object has a time and a PE count for each of the 32 PMTs. From that, the 2D position of the flash, $z\pm \sigma_z$ and $y\pm \sigma_y$, is calculated. These can be compared with the center of deposited charge of all Pandora neutrino candidates. This comparison has the implicit assumption that the light will be emitted in the same relative fraction as the charge is deposited by the final state particles. This is not entirely correct since the amount of scintillation light produced per deposited energy unit is particle dependent. Nevertheless, compared to the coarse resolution of the PMT grid, this approximation is justified.

After applying rectangular cuts to make sure the position of the reconstructed Pandora neutrino candidate is compatible with the reconstructed flash object, it is still likely that an event contains more than one candidate. The single best candidate in each event can be chosen by comparing the shape of the reconstructed flash with the deposited charge in the TPC associated to the reconstructed Pandora candidate.  

\begin{figure}[!htbp]
\centering
\includegraphics[width=0.7\textwidth]{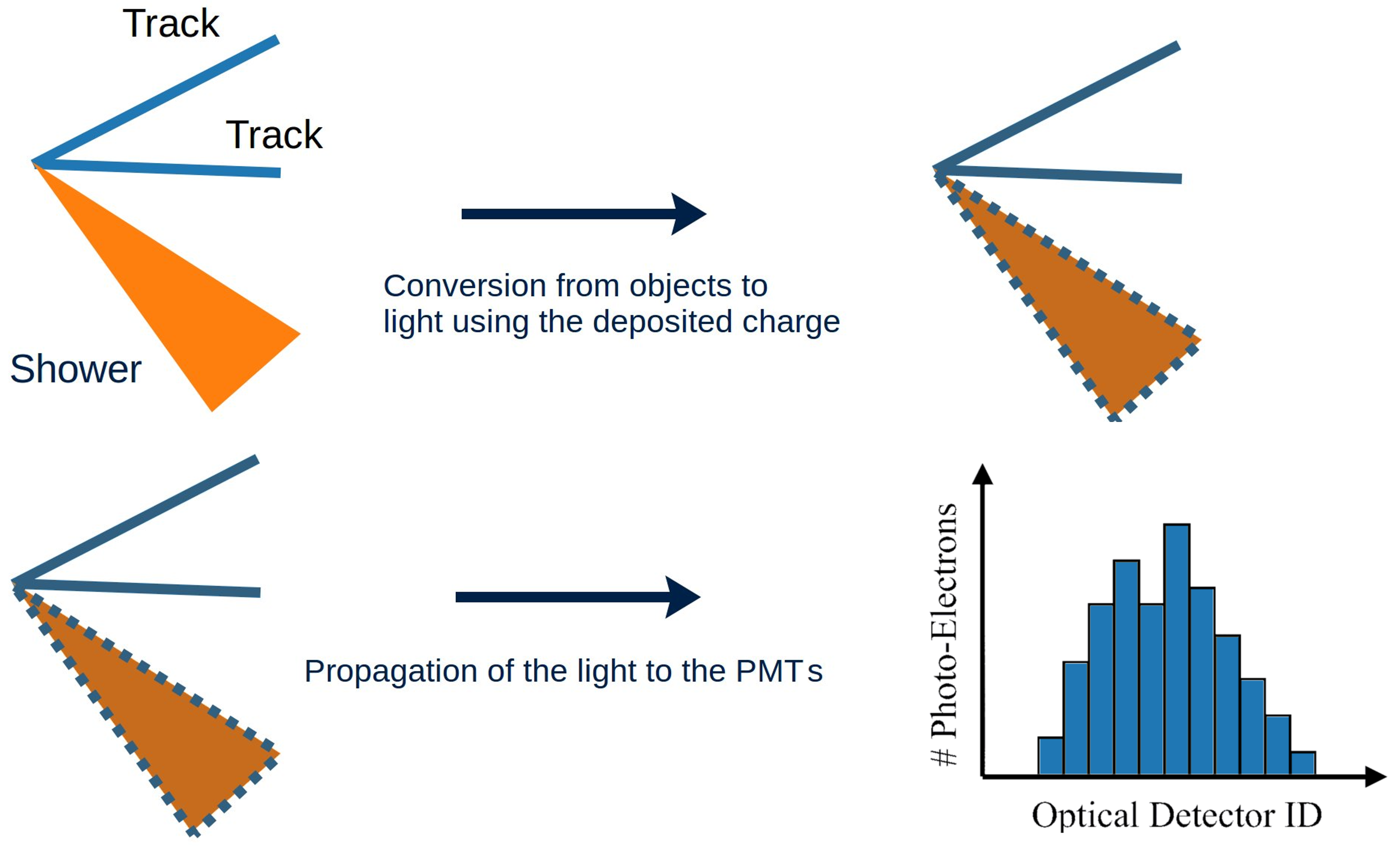} 
\caption{Schematic of the construction of a flash hypothesis for a neutrino candidate.} 
\label{fig:flashdrawing}
\end{figure}

The principle of flash-matching is described in Figure~\ref{fig:flashdrawing}; a flash hypothesis can be constructed for each candidate neutrino interaction using only data recorded by the time projection chamber.
\begin{itemize}
\item for every neutrino candidate, the spatial distribution of deposited charge is collected from the associated showers and tracks;
\item the spatial distribution of the deposited charge is translated into an estimation of the emitted scintillation light. These scintillation photons are then propagated towards the PMTs to construct a flash hypothesis using only time projection chamber information;
\item the flash-matching algorithm compares the reconstructed flash object as seen by the PMT's with the flash hypothesis for all possible neutrino candidates and picks the best matching candidate. For this, a binned likelihood of the PMT spectrum is optimised.
\end{itemize}


\section{Conclusion}

This note explained that TPC information alone is insufficient to discriminate against reconstructed neutrino candidates originating from activity due to cosmic interactions. Therefore, the optical reconstruction algorithms were extended and are shown to be applicable to electron neutrino interactions. Qualitative tests were done to motivate the importance of flash-matching in the Pandora electron neutrino selection chain.


\bibliographystyle{plain}
\bibliography{./bib}

\end{document}